\documentclass[%
aps,
amsmath,amssymb,
reprint,
superscriptaddress
]{revtex4-2}

\usepackage{graphicx}                           
\usepackage{amsfonts, amsmath, amssymb, amsbsy} 
\usepackage{siunitx}
\usepackage{url}                                
\usepackage{textcomp}                           
\usepackage{lipsum}
\usepackage{hyperref}
\hypersetup{
	breaklinks=true,
	colorlinks=true,
	linkcolor=[RGB]{40,42,143},
	filecolor=[RGB]{40,42,143},
	citecolor=[RGB]{40,42,143},      
	urlcolor =[RGB]{40,42,143},
}
\usepackage{mathtools}
\usepackage[normalem]{ulem}

\newcommand{\bx}  {\boldsymbol{x}}
\newcommand{\hlf}{\textstyle\frac{1}{2}}
\newcommand\norm[1]{\left\lVert#1\right\rVert}

\begin{document}
	
	\title[Noncollinear wave interactions]{Nonlinear interaction of two cross-propagating plane waves}
	
	\author{A.~Matalliotakis}
	\altaffiliation{Electronic mail: \url{a.matalliotakis@tudelft.nl}}
	\affiliation{Section of Medical Imaging, Department of Imaging Physics, Faculty of Applied Sciences, Delft University of Technology, 2628 CJ Delft, The Netherlands}
	\author{D.~Maresca}
	\affiliation{Section of Medical Imaging, Department of Imaging Physics, Faculty of Applied Sciences, Delft University of Technology, 2628 CJ Delft, The Netherlands}	
	
	\author{M.D.~Verweij}
	\affiliation{Section of Medical Imaging, Department of Imaging Physics, Faculty of Applied Sciences, Delft University of Technology, 2628 CJ Delft, The Netherlands}
	\affiliation{Section of Biomedical Engineering, Department of Cardiology, Erasmus University Medical Center, 3000 CA Rotterdam, the Netherlands}
	
	
	\begin{abstract} 
		An ideal contrast-enhanced ultrasound image should display microbubble-induced nonlinearities while avoiding wave propagation nonlinearities. One of the most successful ultrasound pulse sequences to disentangle these nonlinear effects relies on the transmission of cross-propagating plane waves. However, theory describing the noncollinear nonlinear interaction of two finite plane waves has not been fully developed and a better understanding of these effects would improve contrast-enhanced ultrasound imaging further. Here, local nonlinear interactions at the intersection of two plane-waves are investigated by extending the Westervelt equation with a term including the Lagrangian density. The Iterative Nonlinear Contrast Source (INCS) method is employed to numerically solve this full nonlinear wave equation for two 3D finite cross-propagating pulsed plane waves. In addition, analytical expressions for the cross-propagation of two infinite continuous plane waves are derived. Numerical results obtained with INCS show good agreement with the analytical expressions. Overall, the generated results show that the pressure associated with local nonlinear effects is two orders of magnitude lower than the pressure associated with global nonlinear effects. Local nonlinear effects are therefore expected to be negligible in the context of single-shot ultrasound imaging, but they may influence approaches that subtract pressure fields such as amplitude modulation or pulse inversion.
	\end{abstract}
	
	\maketitle
	
	%
	\section{Introduction} \label{sec:Introduction}
	%
	\begin{figure*}[tbph!]
		\includegraphics{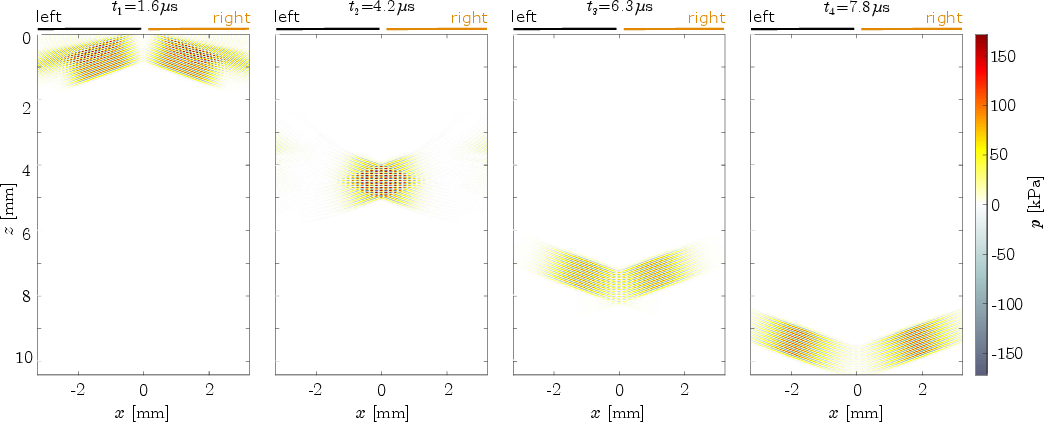}
		\caption{(Color online) Interaction of two cross-propagating beams of finite, pulsed plane waves, at four different time instants: $t_1$ is just before interaction, $t_2$ is when the area of interaction is at a maximum, $t_3$ is just before separation, and $t_4$ is after separation. Transmitting apertures of the ultrasound array are indicated in black and orange at z = 0.} 
		\label{fig:TimeShots}
	\end{figure*}
	The combination of ultrasound echography with lipid-shelled microbubbles administered in the blood stream has enabled quantitative imaging of tissue perfusion\cite{Porter} and super-resolution imaging of the microvasculature in humans.\cite{Demene} Pulse sequences exploit the nonlinear response of resonant microbubbles in an acoustic field to detect their presence in a linearly scattering tissue environment.\cite{Averkiou}. In practice, pulse sequences are prone to nonlinear wave propagation artifacts and tend to missclassify tissue as microbubble domain, even at low acoustic pressure. Renaud et al. \cite{Renaud} and Maresca et al. \cite{xAM} have shown that an amplitude modulation (AM) sequence based on the noncollinear interaction of two ultrasonic wavefronts can significantly reduce such artifacts. A schematic example of the cross-propagation of two finite pulsed plane waves is depicted in Fig.~\ref{fig:TimeShots}. To further improve the specificity of contrast-enhanced ultrasound imaging, a full understanding of nonlinear effects occurring in tissue devoid of ultrasound contrast agents is critical.
	
\begin{figure*}[t!]
		\includegraphics[height=15.5cm]{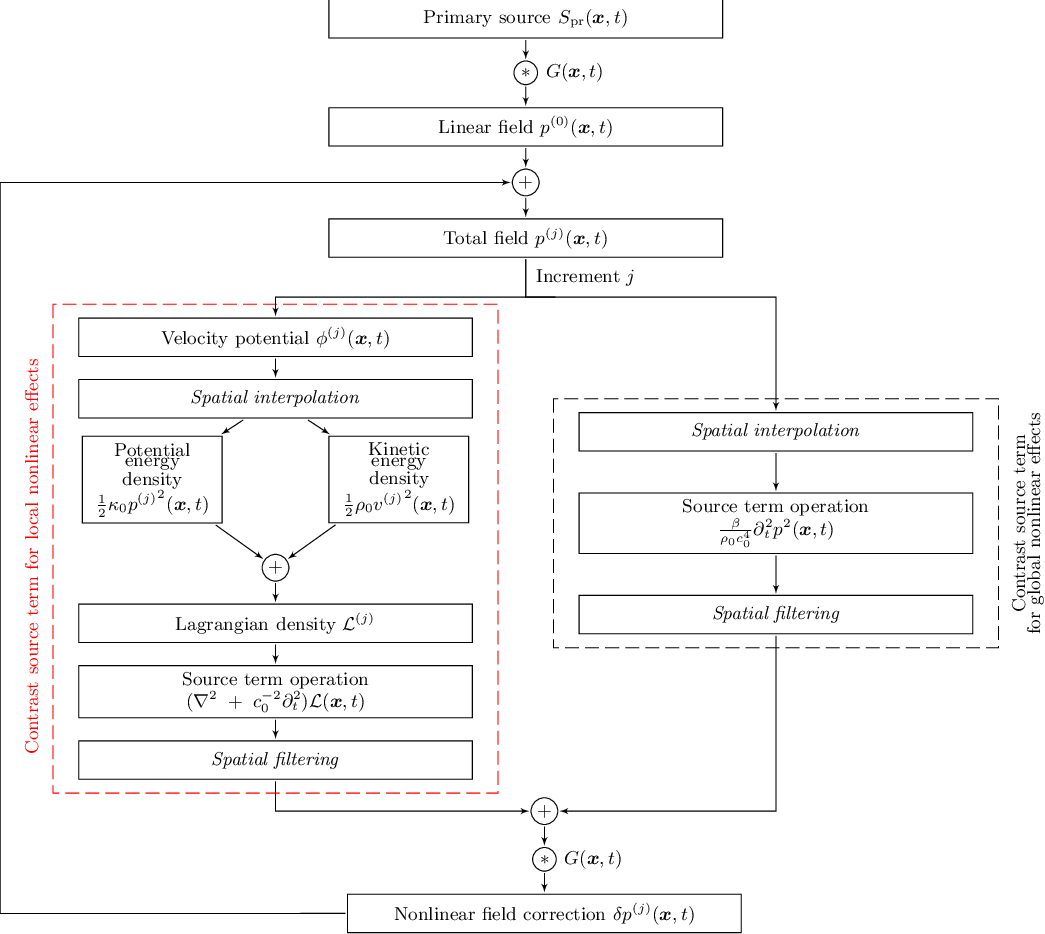}
		\caption{(Color online) Schematic diagram for the INCS method with extension to account for local nonlinear effects (red).} 
		\label{fig:BlockDiagram}
	\end{figure*}

	Previous studies\cite{Aanonsen,Tjottas,Hamilton1,Hamilton2} have shown that local nonlinear effects emerge from the noncollinear interaction of plane waves. Similar observations were reported for parametric acoustic arrays\cite{CervenkaBednariK}. To date, simulation tools for solving the Westervelt equation capture global, i.e. cumulative, nonlinear effects but neglect local nonlinear interactions. Local nonlinear effects manifest where the wave field does not resemble a simple progressive plane wave and the potential energy density of the resulting wave is not equal to its kinetic energy density, which happens at the intersection of two cross propagating plane waves. We will therefore extend the INCS\cite{Koos_Thesis,INCS} method to solve the full nonlinear wave equation accounting for both global and local nonlinear effects. In the original INCS method, global nonlinear effects are accounted for by considering the nonlinear term in the Westervelt equation as a contrast source that acts in a linear background medium. Here, we will introduce an additional contrast source term to account for local nonlinear effects, as explained in Fig.~\ref{fig:BlockDiagram}. INCS computes the acoustic pressure due to a source with a pulsed excitation in a 4D spatiotemporal domain. The directional independence of INCS makes it well adapted to the computation of local nonlinearities generated by the noncollinear interaction of cross-propagating plane waves.
	
	In this article, we will explain how INCS is extended to include local nonlinear medium effects. Initially, the velocity potential is computed from the already calculated pressure field. Next, spatial interpolation is used to upsample the pressure and the velocity potential before computing the potential and kinetic energy densities. These terms then are used to calculate the Lagrangian density. Subsequently, the contrast source term representing the local nonlinear effects is obtained by taking the relevant spatial and temporal derivatives. Spatial filtering is then applied to downsample the spatiotemporal contrast source to its initial grid size. After the convolution with the Green's function of the linear background medium, the nonlinear field correction to the incident pressure field is calculated. Iteration of this approach provides an increasingly accurate solution to the nonlinear wave problem.
	
	The manuscript is organized as follows: fundamentals of INCS are described in Section~\ref{sec:INCS} and cover the addition of a contrast source term accounting for local nonlinear effects. Section~\ref{sec:analytical} derives the analytical expressions for the interaction of two infinite, continuous, cross-propagating plane waves. In Section~\ref{sec:Numerical_Results}, INCS results for two 3D cross-propagating plane waves travelling at a $20^{\circ}$ angle are presented and compared with the analytical derivations. Results for the pressure fields generated by a linear array with hyperbolic time delays are also reported. Conclusions are given in Section~\ref{sec:conclusions}.
	%
	\section{Fundamentals of INCS}\label{sec:INCS}
	%
	\subsection{Linear Field}
	The linear pressure field generated by an external source in a linear, homogeneous acoustic medium is described by the wave equation
	\begin{equation}
		c^{-2}_{0}\frac{\partial^{2} p(\bx,t)}{\partial t}-\nabla^2 p(\bx,t)= S_\mathrm{pr}(\bx,t) ,
		\label{eq:LosslessLinWestervelt}
	\end{equation}
	Here, $\bx$ [m] is the Cartesian position vector, and $t$ [s] is the time.  The symbol $p(\bx,t)$ [Pa] indicates the acoustic pressure, $c_{0}=1/\sqrt{\rho_{0}{\kappa_{0}}}$ [m/s] is the small signal sound speed in the background medium, where $\rho_{0}$ [kg$\cdot \mathrm{m}^{-3}$] is the mass density and ${\kappa_{0}}$ [$\mathrm{Pa}^{-1}$] is the compressibility. The Laplacian operator ${\nabla}^2$ generates the sum of the second order spatial derivatives. The acoustic field is generated by the primary source term
	\begin{equation}
		S_\mathrm{pr}(\bx,t)
		= \rho_{0}\partial_{t}q(\bx,t)
		-\nabla\cdot\boldsymbol{f}(\bx,t).
		\label{eq:PrimarySource}
	\end{equation}
	which contains the volume injection rate density ${q(\bx,t)}$ [$\mathrm{s}^{-1}$] and the volume force density $\boldsymbol{f}(\bx,t)$ [N $\mathrm{m}^{-3}$]. Pressure jump conditions for the velocity or the pressure can be used to represent a source with a plane aperture, e.g. a phased array transducer.
	
	It is convenient to denote the method that solves the linear wave equation for a given source by the operator $\mathcal{G}$. The field $p$ due to a source $S$ can then be obtained as
	\begin{align}
		p(\bx,t) &=\mathcal{G}[S] \nonumber\\
		&=\int_{\mathcal{T}}\int_{\mathcal{X}}S(\bx',t)\,G(\bx-\bx',t-t')\,\bx'dt'.
		\label{eq:LinearFieldEstimate}
	\end{align}
	In this equation, the Green's function $G(\bx,t)$ is the solution of Eq.~(\ref{eq:LosslessLinWestervelt}) for a spatiotemporal impulsive source, which is known analytically, and the convolution takes place over the spatial domain $\mathcal{X}$ and over the temporal domain $\mathcal{T}$ of the source $S$.
	
	In particular, the acoustic pressure $p^{(0)}=\mathcal{G}[S_\mathrm{pr}]$ is the linear field of the primary source. This field will act as the zeroth order iteration for the Neumann iterative scheme that will be used to solve the nonlinear problem.
	\subsection{Nonlinear field due to global nonlinear effects}
	In medical ultrasound, the nonlinear behaviour of the medium can have a significant impact on the propagation of the acoustic signals. If we account for nonlinear terms up to second order in the acoustic quantities and assume that the cumulative nonlinear behavior dominates the local behavior\cite{Aanonsen,Tjottas}, then it is sufficient to extend Eq.~(\ref{eq:LosslessLinWestervelt}) to the Westervelt equation
	\begin{equation}
		{c^{-2}_{0}}\frac{\partial^{2} p}{\partial t^2}-\nabla^{2}p= S_\mathrm{pr} + S_\mathrm{nl}.
		\label{eq:LosslessWestervelt}
	\end{equation}
	The additional nonlinear term is given by
	\begin{equation}
		S_\mathrm{nl}(p)=\frac{\beta}{\rho_{0}c^{4}_{0}}\partial^{2}_{t}p^{2},
		\label{eq:NonlinearCSTerm}
	\end{equation}
	in which $\beta$ is the coefficient of nonlinearity of the medium. In INCS, this term acts as a contrast source term that accounts for the global nonlinear behavior of the medium. A first order nonlinear correction $\delta p^{(1)}=\mathcal{G}[S_\mathrm{nl}(p^{(0)})]$ to the acoustic pressure field $p^{(0)}$ can be calculated by following the same approach as in Eq.~(\ref{eq:LinearFieldEstimate}), where the integration now takes place over the spatial domain $\mathcal{X}_{\mathrm{nl}}$ and the temporal domain $\mathcal{T}_{\mathrm{nl}}$ of the nonlinear contrast source. The result is the first-order corrected field $p^{(1)}=p^{(0)}+\delta p^{(1)}$. The correction procedure can be repeated by substituting the corrected field in the nonlinear contrast source $S_\mathrm{nl}(p)$ and computing an improved nonlinear correction $\delta p$ to the acoustic pressure field $p^{(0)}$. This leads to a Neumann iterative scheme that is given by the following set of equations
	\begin{align}
		p^{(0)} &= \mathcal{G}[S_\mathrm{pr}],\label{eq:NeumannIterative_0}\\
		p^{(j)} &= p^{(0)} + \mathcal{G}[S_\mathrm{nl}(p^{(j-1)})], \quad\text{if}\;j\geq 1.
		\label{eq:NeumannIterative_j}
	\end{align}
	Within this scheme, other contrast sources can be accommodated that represent attenuation\cite{Attenuative_media1,Attenuative_media2,Libe,Lossy_Green}, inhomonegeous medium properties\cite{INCS_Inhom}, or a population of scatterers such as nonlinear oscillating microbbubles.\cite{Bubble_Cloud}
	\subsection{Nonlinear field  due to local nonlinear effects}
	We can extend Eq.~(\ref{eq:LosslessWestervelt}) by introducing a contrast source term $S_\mathcal{L}(\bx,t)$ that represents local nonlinear effects. We then get 
	\begin{equation}
		{c^{-2}_{0}}\frac{\partial^{2} p(\bx,t)}{\partial t}-\nabla^{2}p(\bx,t)= S_\mathrm{pr}(\bx,t) + S_\mathrm{nl}(\bx,t) + S_\mathcal{L}(\bx,t),
		\label{eq:FullNonlinearWaveEquation}
	\end{equation}
	where $S_\mathcal{L}(\bx,t)$ is described by
	\begin{equation}
		S_\mathcal{L}(\bx,t)=(\nabla^{2} + {c^{-2}_{0}}\partial^{2}_{t}) \mathcal{L}(\bx,t).
		\label{eq:LagrangianSourceTerm}
	\end{equation}
	Here, $\mathcal{L}(\bx,t)$ is the Lagrangian density 
	\begin{equation}
		\mathcal{L}(\bx,t)=\hlf\rho_0\norm{\boldsymbol{u}(\bx,t)}^2 - \hlf\kappa_0 p^{2}(\bx,t),
		\label{eq:LagrangianDensity}
	\end{equation}
	where $\boldsymbol{u}(\bx,t)$ is the particle velocity. The Lagrangian density is the difference between the kinetic energy density and the potential energy density of the acoustic wave. For a plane wave, the Lagrangian density equals zero, but this is not the case for two noncollinear interacting plane waves.
	
	In the current framework, we need an expression that gives the particle velocity as a function of pressure. This is achieved through the velocity potential $\phi(\bx,t)$, which is defined as
	\begin{equation}
		\boldsymbol{u}(\bx,t)=\nabla \phi(\bx,t). 
		\label{eq:AcousticPotentialEq2}
	\end{equation}
	In first order, the relation between $p(\bx,t)$ and $\phi(\bx,t)$ then becomes
	\begin{equation}
		p(\bx,t)=-\rho_{0}\frac{\partial\phi(\bx,t)}{\partial t}.
		\label{eq:AcousticPotentialEq1}
	\end{equation}
	In this equation, terms of order two and higher in the acoustic quantities are neglected. This results in neglecting terms of order three and higher in $p^2(\bx,t)$ and thus in $\mathcal{L}(\bx,t)$, which is allowed because the wave equation in Eq.~(\ref{eq:FullNonlinearWaveEquation}) is only accurate up till order two in the acoustic quantities. With Eqs.~(\ref{eq:AcousticPotentialEq2}) and (\ref{eq:AcousticPotentialEq1}), Eq.(\ref{eq:LagrangianDensity}) can be rewritten as
	\begin{equation}
		\mathcal{L}(\bx,t)=\frac{\rho_{0}}{2}\left\{\norm{\nabla\phi(\bx,t)}^2 - c^{-2}_0\left[\frac{\partial\phi(\bx,t)}{\partial t}\right]^2\right\}.
		\label{eq:LagrangianDensityphi}
	\end{equation}
	Up till second order accuracy, this is identical to\cite{Aanonsen}
	\begin{equation}
		\mathcal{L}(\bx,t)=\frac{\rho_{0}}{4}\left(\nabla^2 - c^{-2}_0\frac{\partial^2}{\partial t^2}\right)\,\phi^2(\bx,t).
		\label{eq:LagrangianDensityphi_SecondOrder}
	\end{equation}
	%
	\section{Analytical expressions}\label{sec:analytical}
	\subsection{Pressure and velocity potential}	
	For comparison and benchmarking purposes, here we will derive analytical expressions for the Lagrangian density and the resulting contrast source for cross-propagating plane waves. Specifically, we consider two infinite, steady-state, plane acoustic waves that are propagating in a homogeneous medium. Each wave has an angular frequency $\omega$ and propagates under an angle $\pm\theta$ with the $z$-axis, respectively. The medium has a wave speed $c_{0}$, a density of mass $\rho_0$ and a coefficient of nonlinearity $\beta$. Without loss of generality, we consider the plane $y=0$ and omit the variable $y$. In this case, the total incident pressure $p(x,z,t)$ may be written as
	\begin{equation}
		p=P_0
		[\sin(\omega t-k_x x - k_z z ) + \sin(\omega t + k_x x - k_z z )],
		\label{eq:PressureAnalyticalCrossPropagating}
	\end{equation} 
	where
	\begin{equation}
		k_x = \frac{\omega}{c_0} \sin(\theta) =k \sin(\theta),
		\label{eq:k_x}
	\end{equation}
	\begin{equation}
		k_z = \frac{\omega}{c_0}\cos(\theta) = k \cos(\theta).
		\label{eq:k_z}
	\end{equation}
	Using Eq.~(\ref{eq:AcousticPotentialEq1}), we find
	\begin{equation}
		\phi(x,z,t)=\frac{P_0}{\rho_0 \omega}[\cos(\omega t - k_x x -k_z z )+\cos(\omega t + k_x x -k_z z )].
		\label{eq:phi_planewaves}
	\end{equation}
	\subsection{Lagrangian Density}
	To compute the Lagrangian density, we first need to compute the square of the velocity potential, which is
	\begin{equation}
		\phi^2 = \frac{P^2_0}{\rho^2_0 \omega^2}[I_0 + I_1 + I_2 + I_3 + I_4],
		\label{eq:phi_sq_terms}
	\end{equation}
	where
	\begin{align}
		I_0 &= 1 \label{eq:I0}\\
		I_1 &= \hlf\cos(2\omega t - 2k_x x -2k_z z ),\label{eq:I1}\\
		I_2 &= \hlf\cos(2\omega t + 2k_x x -2k_z z ),\label{eq:I2}\\
		I_3 &= \cos(2\omega t - 2k_z z ),\label{eq:I3}\\
		I_4 &= \cos(2k_x x).\label{eq:I4}
	\end{align}
	In order to derive this expression from Eq.~(\ref{eq:phi_planewaves}), we have used the trigonometric identity
	\begin{equation}
		\cos(a+b) + \cos(a-b)=2\cos(a)\cos(b).
		\label{eq:addcos}
	\end{equation}
	When computing the Lagrangian density, we can beforehand discard some terms in Eq.~(\ref{eq:phi_sq_terms}). Because $I_0$ is a constant, it will not contribute to $\mathcal{L}$. Furthermore, $I_1$ and $I_2$ are a representation of plane waves, which also do not contribute to $\mathcal{L}$. Therefore, we can continue the derivation with the remaining terms, giving
	\begin{align}
		\mathcal{L}  &= \frac{P^2_0}{4\rho_0\omega^2} \left( \nabla^2-\frac{1}{c^2_0}\frac{\partial^2}{\partial t^2} \right) (I_3+I_4) \nonumber\\
		&= \frac{P^2_0}{4\rho_0\omega^2} \left( \nabla^2-\frac{1}{c^2_0}\frac{\partial^2}{\partial t^2} \right) \left[\cos(2\omega t-2k_zz)+\cos(2k_xx)\right] \nonumber\\
		&=\frac{P^2_0}{\rho_0 c^2_0}\sin^2(\theta)[\cos(2\omega t - 2 k_z z ) - \cos(2 k_x x)],\label{eq:Lagrangian_analytical}
	\end{align}
	where we have used
	\begin{equation}
		\frac{\omega^2}{c^2_0}-k^2_z=k^2_x.
		\label{eq:wavenumber}
	\end{equation}
	\begin{figure}[b!]
		\includegraphics{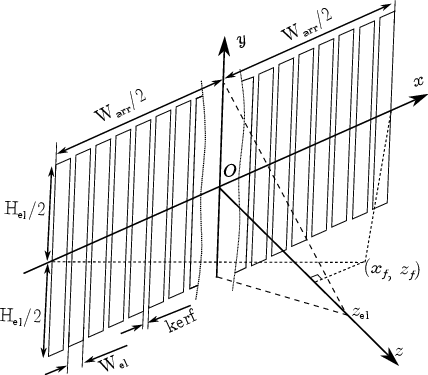}
		\caption{Sketch of the geometry of the phased array that generates the incident field. } 
		\label{fig:Transducer}
	\end{figure}
	\subsection{\label{subsec:sl}Contrast source term for local nonlinear effects}
	Here, we derive the analytical expression for the contrast source in Eq.~(\ref{eq:LagrangianSourceTerm}). The first term equals 
	\begin{align}
		\nabla^2 \mathcal{L} =& -4\frac{P^2_0 \omega^2}{\rho_0 c^4_0}\sin^2(\theta)\,[\cos^2(\theta) \cos(2\omega t -2k_z z)
		\nonumber\\
		&\hspace{30mm}- \sin^2(\theta)  \cos(2 k_x x)],
		\label{eq:nabla_squared}
	\end{align}
	and the second term is
	\begin{equation}
		c^{-2}_0 \frac{\partial^2 \mathcal{L}}{\partial t^2} =-4\frac{P^2_0\omega^2}{\rho_0 c^4_0} \sin^2(\theta)\cos(2\omega t -2k_z z).
		\label{eq:second_time_deriv}
	\end{equation}
	By adding Eqs.~(\ref{eq:nabla_squared}) and  (\ref{eq:second_time_deriv}), the source term for the local nonlinear effects is obtained as
	\begin{figure*}[bpt!]
		\includegraphics{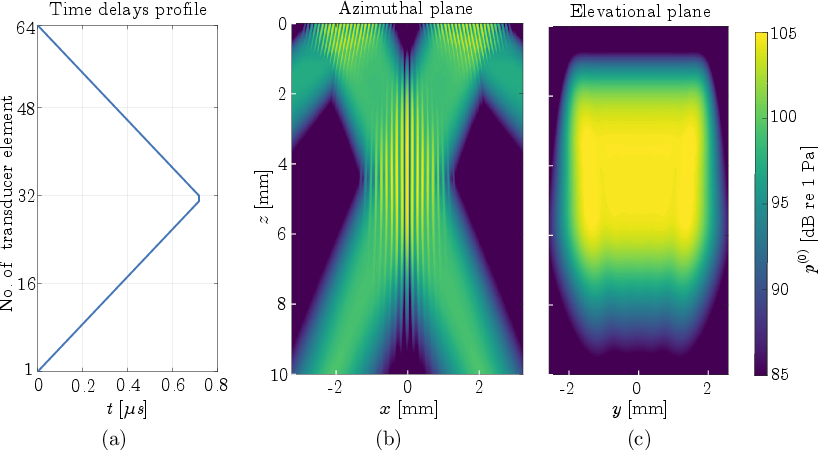}
		\caption{\label{fig:PhasedLinearPressureField}(Color online) The incident pressure field for cross-propagating beams generated by the phased array. (a) Delay profile for generating two beams with plane wave fronts that propagate under angles $\theta=\pm 20^{\circ}$ with respect to the $z$-axis. (b) Maximum pressure profiles in the azimuthal plane $y=0\;\mathrm{mm}$. (c) Maximum pressure profiles in the elevational plane  $x=0\;\mathrm{mm}$.}
	\end{figure*}
	\begin{equation}
		\begin{split}
			S_{\mathcal{L}} =&-8\frac{P^2_0 \omega^2 }{\rho_0 c^4_0}\sin^2(\theta)\cos(2\omega t -2k_z z) \\
			&+4\frac{P^2_0 \omega^2 }{\rho_0 c^4_0}\sin^4(\theta)[\cos(2\omega t -2k_z z)+\cos(2 k_x x)],
		\end{split}
		\label{eq:ST_LocalNL}
	\end{equation}
	where we have used
	\begin{equation}
		1+\cos^2(a) = 2 - \sin^2(a).
		\label{eq:cossquare}
	\end{equation}
	\subsection{Source term for global nonlinear effects}
	To find the analytical expression for the contrast source in Eq.~(\ref{eq:NonlinearCSTerm}), we first compute the square of the acoustic pressure, which is
	\begin{equation}
		p^2 = P^2_0[J_0+J_1+J_2+J_3+J_4],
		\label{eq:p_sq_terms}
	\end{equation}
	where
	\begin{align}
		J_0 &= 1 \label{eq:J0}\\
		J_1 &= -\hlf\cos(2\omega t - 2k_x x -2k_z z ),\label{eq:J1}\\
		J_2 &= -\hlf\cos(2\omega t + 2k_x x -2k_z z ),\label{eq:J2}\\
		J_3 &= -\cos(2\omega t - 2k_z z ),\label{eq:J3}\\
		J_4 &= \cos(2k_x x),\label{eq:J4}
	\end{align}
	in which we have used
	\begin{equation}
		-\cos(a+b) + \cos(a-b)=2\sin(a)\sin(b).
		\label{eq:minuscosa+b}
	\end{equation}
	Because $J_0$ is a constant, and $J
	_4$ is independent of time, these terms will have no contribution to $S_\mathrm{nl}$. Therefore, we can continue the derivation with the remaining terms, giving
	\begin{align}
		S_\mathrm{nl} =& \frac{\beta P^2_0}{\rho_0 c^4_0}\frac{\partial^2}{\partial t^2}(J_1+J_2+J_3) \nonumber\\
		=&8\frac{\beta P^2_0 \omega^2}{\rho_0 c^4_0}\cos^2(k_x x) \cos(2\omega t -2k_z z).
		\label{eq:ST_MediumNL}
	\end{align}
	%
	\section{NUMERICAL RESULTS}\label{sec:Numerical_Results}
	%
	\subsection{Configuration}\label{subsec:configuration}
	After the derivation of the analytical expressions, we can continue with the comparison of the numerical outcomes obtained by INCS with the results derived in Sec.~\ref{sec:analytical}. In the current section, we consider a computational domain of dimensions $X\times Y\times Z=6.4\;\mathrm{mm}\times 5\;\mathrm{mm}\times 10\;\mathrm{mm}$. The medium considered in the analysis is water, characterized by a density of mass of $\rho = 1060\;\mathrm{kg/m^3}$, and a speed of sound $c_0 = 1482\;\mathrm{m/s}$.
	
	The incident beam has a center frequency $f_0=15\;\mathrm{MHz}$ and is produced by a phased array transducer comprising of 64 elements, each with dimensions $H_\mathrm{el}\times W_\mathrm{el}=5\;\mathrm{mm}\times0.1\;\mathrm{mm}$, and a kerf with zero width. This configuration results in an aperture with a width of $W_\mathrm{ap}= 6.4\;\mathrm{mm}$. The selection of this particular size was done to shift the natural focus of the aperture out of the computational domain, to obtain two cross-propagating pressure fields that closely approximate plane waves. Additionally, the origin of the coordinate system has been positioned at the center of the transducer aperture. A sketch depicting the geometry of the phased array is presented in Fig.~\ref{fig:Transducer}.
	
	The time-varying pressure at the surface of the elements is given by the expression
	\begin{equation}
		p(t)=P_0\,\mathrm{exp}\left[-\left(\frac{t-T_d}{T_w/2}\right)^2\right]\mathrm{sin}[2\pi f_0 (t-T_d)],
		\label{eq:Time_Signature}
	\end{equation}
	where we have chosen $T_w=10/f_0$, which represents the duration of the Gaussian envelope, and $T_d=10/f_0+\Delta_n$, which is the total time delay. The latter consists of a fixed delay for keeping $p(t)\approx 0$ at $t=0$, plus a delay per element for the beam steering $\Delta_n$. The maximum surface pressure of the elements is $P_0=100\;\mathrm{kPa}$.
	
	For accurately solving the full nonlinear wave equation up to the second harmonic, a sampling frequency of 96 MHz has been used to discretize the spatiotemporal domain. To reduce the artifacts arising from the edges of the aperture, a Tukey apodization with a cosine fraction of 0.7 has been applied.
	\begin{figure}[t!]
		\includegraphics{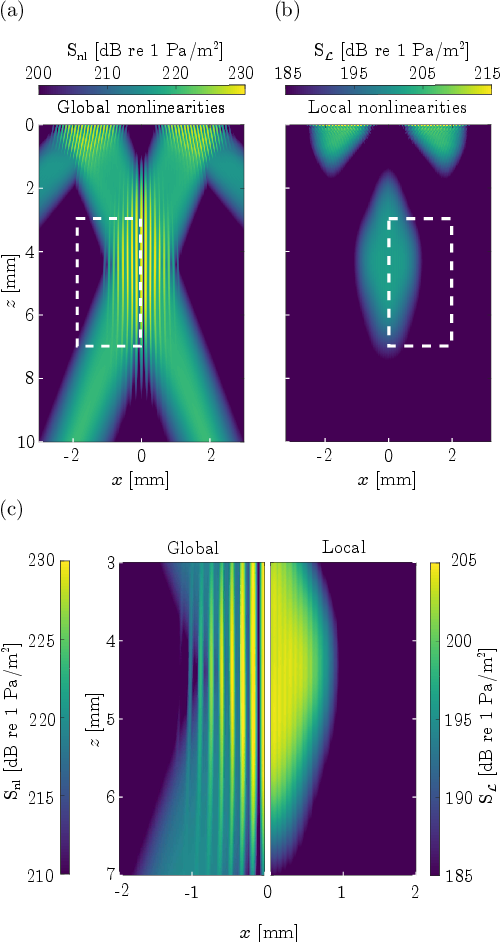}
		\caption{(Color online) Maximum values of the nonlinear contrast source terms in the azimuthal plane $y=0\;\mathrm{mm}$ for the beams generated by the phased array. (a) Source term for global nonlinear effects. (b) Source term for local nonlinear effects. (c) Detailed view of both source terms side by side. }		 
		\label{fig:SourceTerms_BeamProfiles}
	\end{figure}
	\subsection{Crossing beams}\label{sec:crossing_beams}
	First we will show results for two cross-propagating beams with finite plane wave fronts. To generate two plane waves that propagate under angles of $-20^{\circ}$ and $20^{\circ}$, respectively, a triangular time delay profile is applied as depicted in Fig.~\ref{fig:PhasedLinearPressureField}(a). The incident beams in the azimuthal plane $y=0\;\mathrm{mm}$ and the elevational plane $x=0\;\mathrm{mm}$ are shown in Figs.~\ref{fig:PhasedLinearPressureField}(b) and (c), respectively. In the azimuthal plane, the beams show a constant lateral width. The emitted pulses are long enough to resemble a steady-state wave. In this way, we can compare the results of INCS with the ones from the analytical expressions. Moreover, the maximum pressure in the intersection area is $181\;\mathrm{kPa}$, which corresponds to a mechanical index of 0.047. Although we use a low mechanical index, the values for local and global nonlinear effects will be scaling similarly as they are both a function of the square of the pressure. 
	\begin{figure}[b!]
		\includegraphics{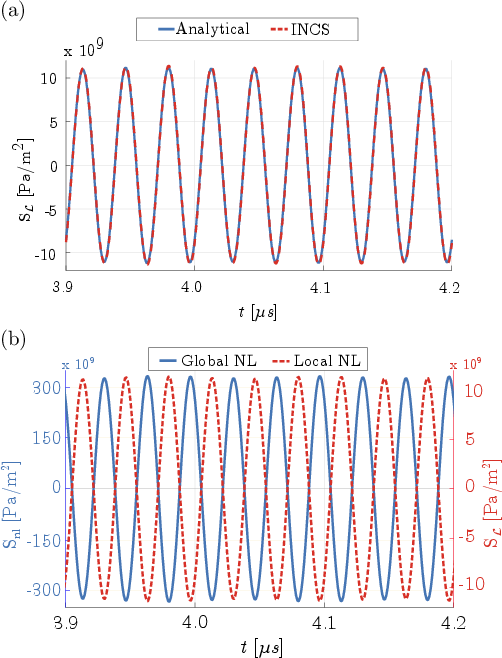}
		\caption{(Color online) Time domain signatures of the contrast sources at the center of intersection $(x, y, z)=(0\;\mathrm{mm}, 0\;\mathrm{mm}, 4.3\;\mathrm{mm})$. (a) Analytical results (blue, continuous) and numerical INCS results (red, dashed) for the contrast source for local nonlinear effects. (b) Numerical INCS results for the contrast source terms for global nonlinear effects (blue, continuous) and local nonlinear effects (red, dashed).} 
		\label{fig:SourceTerms_TimeSignatures}
	\end{figure}
	\begin{figure*}[tb!]
		\includegraphics{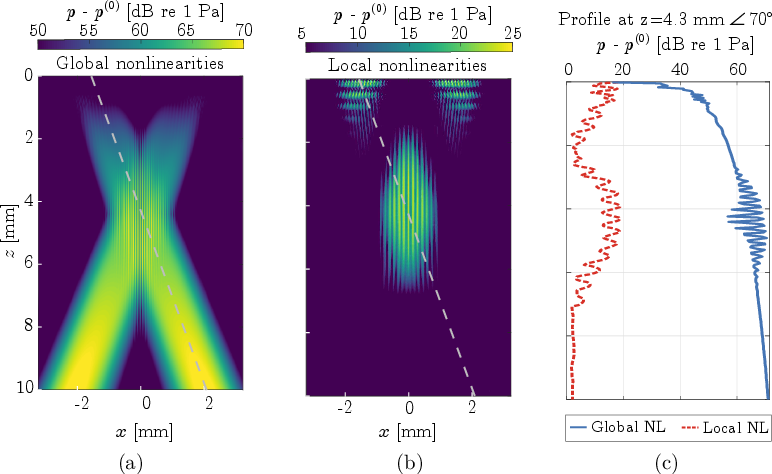}
		\caption{(Color online) Maximum pressure profiles in the azimuthal plane $y=0\;\mathrm{mm}$ for the beams generated by the phased array. (a) Pressure due to global nonlinear effects. (b) Pressure due to local nonlinear effects. (c) Maximum values of the pressures in (a) and (b) at the centerline of the beam emitted by the left part of the phased array.} 
		\label{fig:Pressure_Local_MediumNL}
	\end{figure*}

	In Fig.~\ref{fig:SourceTerms_BeamProfiles}(a) and (b), the respective magnitudes of the source terms for global and local nonlinear effects are depicted. The magnitude of the source term for the global nonlinear effects is higher at locations where the incident pressure in Fig.~\ref{fig:PhasedLinearPressureField} is higher. This is because the considered source term depends on the square of the total pressure, which is dominated by the incident pressure. On the other hand, the source term for the local nonlinear effects is stronger at locations where in Fig.~\ref{fig:PhasedLinearPressureField} two waves are crossing under an angle, i.e. at the intersection of the two main beams, and also at the intersection of a main beam and its grating lobe. This is because at these locations the kinetic energy density is not equal to the potential energy density. In the regions where there is only a single plane wave, the value of the source term is virtually zero.  The source term for the local nonlinear effects is in general an order of magnitude smaller than the source term for the global nonlinear effects. This is also predicted by the analytical expressions of Sec.~\ref{sec:analytical}. Besides the difference in amplitude, the spatial behavior of the source terms also differs, as demonstrated in Fig.~\ref{fig:SourceTerms_BeamProfiles}(c). In the left plot, a horizontal modulation of the source term for the global nonlinear effects shows up. This is due to the interference of the pressure waves of both beams, which results in a horizontal modulation of the $p^2$ term in the source term, in agreement with Eq.~(\ref{eq:ST_MediumNL}). In the right plot, the horizontal modulation of the source term for the local nonlinear effects is much weaker. This is because the potential energy density depends on $p^2$, while the kinetic energy density term depends on $v^2$. Since both terms spatially alternate in magnitude, the source term will show less horizontal modulation, which agrees with Eq.~(\ref{eq:ST_LocalNL}). The scaling of the colorbars for both plots are chosen to ease the comparison.
	
	The temporal signatures of the analytical and numerical results for the contrast source that represents the local nonlinear effects are presented in Fig.~\ref{fig:SourceTerms_TimeSignatures}(a). The results apply to the point $(x, y, z)=(0\;\mathrm{mm}, 0\;\mathrm{mm}, 4.3\;\mathrm{mm})$, which is the center of intersection of the two plane waves. The depicted time interval spans 9 periods around the center of the incident pressure pulse, to avoid transient effects from the beginning and end of this pulse and to allow comparison with our steady-state analytical results. There is excellent agreement between the analytical results and the numerical results generated by INCS. In  Fig.~\ref{fig:SourceTerms_TimeSignatures}(b), a comparison between the numerically obtained temporal signatures of the contrast source terms that represent the global and local nonlinear effects is presented. The signatures are $180^\circ$ out of phase, which agrees with the fact that the first and dominant term in Eq.~(\ref{eq:ST_LocalNL}) has a minus sign while the expression in Eq.~(\ref{eq:ST_MediumNL}) has not. The peak amplitude of the source term attributed to local nonlinearities is $11.3\times 10^9\;\mathrm{Pa/m}^2$ and the peak amplitude of the source term attributed to global nonlinearities is $330\times 10^9\;\mathrm{Pa/m}^2$. This corresponds to an amplitude rate of approximately 29.2. In the current case with $\theta=20^\circ$ and $\beta=3.21$, the same result is obtained upon comparing Eq.~(\ref{eq:ST_MediumNL}) with Eq.~(\ref{eq:ST_LocalNL}).
	
	Figure \ref{fig:Pressure_Local_MediumNL} shows the pressures that are generated by the source terms in Fig.~\ref{fig:SourceTerms_BeamProfiles}. In panel (a), we see that the global nonlinear effects accumulate along each beam because these effects cause progressive pressure waves. In contrast, we observe in panel (b) that the local nonlinear effects emerge only at locations where the source term is present, but these do not cause progressive pressure waves that propagate into other regions. Although the source term for global nonlinear effects is about 30 times larger than that the source term for local nonlinear effects, the cumulative effect makes that the pressure field of the former is about 500 times larger than the latter.
	\subsection{Focused beam}\label{sec:Focused_phased_array}
	%
	\begin{figure*}[htb!]
		\includegraphics{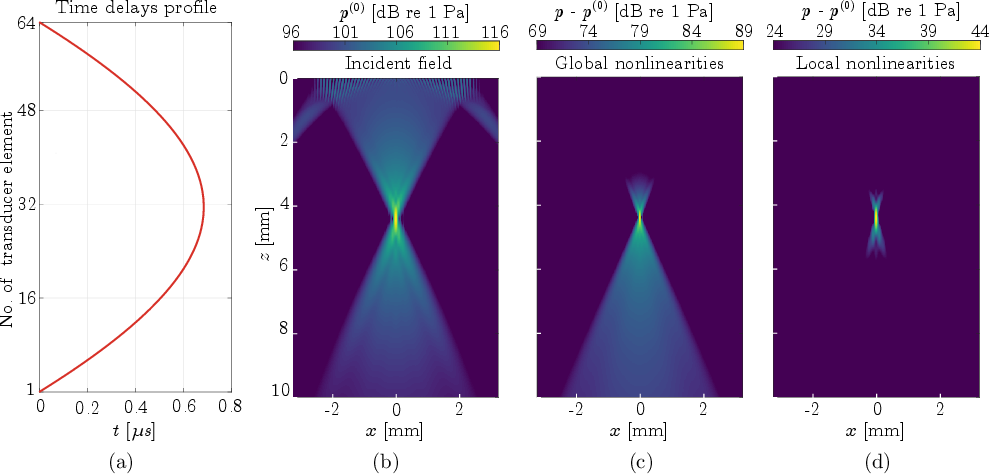}
		\caption{(Color online) Results for a focused beam. (a) Delay profile for generating a single beam focused at $(x, y, z)=(0\;\mathrm{mm}, 0\;\mathrm{mm}, 4.3\;\mathrm{mm})$. (b) Maximum incident pressure generated by the phased array.  (c) Maximum pressure due to global nonlinear effects. (d) Maximum pressure due to local nonlinear effects.} 
		\label{fig:pAM_results}
	\end{figure*}
	Next we will show the results for a focused phased array generating a single focused beam, which is commonly utilized on conventional ultrasound machines for multiple medical applications. The computational domain, geometry of the phased array, and the transmitted pulse are identical to those in Sec.~\ref{sec:Numerical_Results}. The only difference is the time delay profile, which is presented in Fig.~\ref{fig:pAM_results}(a). The time delays are chosen such that the focus is at the same position as the center of the interaction zone of the crossing beams. 
	
	Figure \ref{fig:pAM_results}(b) displays the incident pressure field generated by the focused phased array. Compared to the crossing beams in Fig.~\ref{fig:PhasedLinearPressureField}(a), the maximum amplitude is roughly four times larger at the focus while the length of the focal area is considerably smaller than the interaction zone. The pressure field due to global nonlinear effects is presented in Fig.~\ref{fig:pAM_results}(c). The cumulative behavior of the global nonlinear effects is evident. Figure \ref{fig:pAM_results}(d) shows the pressure field due to local nonlinear effects. In comparison to the global nonlinear effects, these effects are solely occurring  near the focus, i.e. where the waves emitted by the transducer in different directions, interfere constructively. In the current case, the peak amplitude of the global nonlinear effects is about 150 times larger than the peak amplitude of the local nonlinear effects. This is in agreement with the previous findings for focused beams.\cite{Focused_Beam}
	%
	\section{Conclusions}\label{sec:conclusions}
	%
	The Iterative Nonlinear Contrast Source (INCS) method has been extended with an additional contrast source term to simulate local nonlinear effects that are not included in the commonly employed Westervelt equation for nonlinear wave propagation. For two plane waves propagating under an angle, it has been shown both analytically and numerically that this contrast source is nonzero in the interaction zone of the waves. Numerical results show that the pressure associated with local nonlinear effects does not propagate, and indeed is a local phenomenon. This behavior sets local nonlinear effects apart from global nonlinear effects, which cause harmonic pressure waves that accumulate along the entire propagation path of a single plane wave. Moreover, analytical and numerical results for two beams crossing under $20^\circ$ in water show that the pressure associated with nonlinear effects is about two orders of magnitude lower than the pressure associated with global nonlinear effects. The same is observed for a highly focused beam with $f_\# = 1$. Thus, it is expected that local nonlinear effects will not have a significant influence on ultrasound imaging protocols based on single pulse transmissions, but these may have an influence on protocols in which results from multiple pulse transmissions are subtracted, like pulse inversion and amplitude modulation. %
	\begin{acknowledgments}
		This research was supported by the project "Optoacoustic sensor and ultrasonic microbubbles for dosimetry in proton therapy" of the Dutch National Research Agenda which is partly financed by the Dutch Research Council (NWO). We thank G. Renaud for the fruitful discussions, feedback and advice. 
	\end{acknowledgments}
	\section*{Author Declarations}
	The authors have no conflict of interest to disclose.
	
	\section*{Data availability}
	The data that support the findings of this study are available from the corresponding author upon reasonable request.
	
	
\end{document}